\documentclass[twocolumn,prl,showpacs,floatfix]{revtex4}

\usepackage{graphicx}

\begin{document}

\title
{
Transmission and Conversion of Energy by Coupled Soft Gears
}
    
\author
{ 
Akinori Awazu
}

\affiliation
{
Department of Physics, University of Tokyo, Hongou 7-3-1, Bunkyou-ku, Tokyo 113-0033, Japan.
}

\date{\today}

\begin{abstract}
Dynamical aspects of coupled deformable gears are investigated to clarify 
the differences of mechanical properties between the machines consist of 
hard materials and those of soft materials. In particular, the performances 
of two functions, the transmission and the conversion of the energy, are 
compared between the hard and soft gears systems. First, the responses of 
the coupled gears against a constant torque working on one of gears are 
focused for two types of couplings; P) a pair gears are coupled, and 
T) three gears are coupled with forming a regular triangle. In systems with 
the coupling P), we obtain trivial results that the rotational energy can be 
transmitted to other gear only if these gears are hard enough. On the other 
hand, in systems with the coupling T), the transmission of the rotational 
energy to one of the other gears appears only if these gears are soft enough. 
Second, we show the responses of this system in which 
one of gears have contact with a high temperature heat bath and the other 
gears have contact with a $0$ temperature heat bath.
With the coupling T), the directional rotations appear in two gears with 
$0$ temperature heat bath.
Here, the direction of these rotations change depending on the noise 
strength.   
\end{abstract}


\maketitle

Recently, the practical applications of micro- or nano-scale machines are 
hoped by several area of sciences in particular the biology and medical 
science. In the environments where micro or nano scale machines work, the 
influences of the thermal fluctuations is so large that it cannot be 
neglected. In our familiar macro-scale machines, for example, a hard gear 
(rotator) has been known as a useful component to realize several functions. 
In order to get high performance of functions from the populations of hard 
gears, however, each pair of these gears must be always kept closely with 
each other to be coupled tightly. Thus, such machines seem weak for 
fluctuations.
 
On the other hand in biological systems, some kinds of functions for example 
the transmissions or conversions of energy, materials and signals are 
realized efficiently by many types of molecular machines under the large 
influences of thermal fluctuations\cite{cell,motor}. It is remarkable that, 
differing from our familiar parts of macro-scale machines, they 
are generally soft and easy to deform. From these facts, a question arises: 
how different are the mechanical properties and performances between hard 
machines as our familiar macro-scale machines and soft machines as 
molecular machines to realize similar functions? 

In this paper, we demonstrate the simulations of the systems consisting of 
hard gears (rotators) and those of soft gears (rotators) under the same 
situations in order to compare the differences of the mechanical properties 
directly between the machines with hard elements and those with soft 
elements. In particular, we focus on the transmission and the conversion 
of energy realized by coupled gears in the following two situations; 
P) a pair gears are coupled, and T) three gears are put where the centers of 
them form a regular triangle (as Gear 1, Gear 2 and Gear 3 in Fig. 1(b) and 
(d).). The population of some types of rotators 
are focused to understand the dynamical and statistical 
properties of soft materials like the colloidal systems with electric of 
magnetic moment (ferrofluid)\cite{colloid1,colloid2,colloid3,colloid4}, 
recently. Thus, the study of the coupled gears (rotators) has the large 
importance in wide areas of material sciences.

For the system consists of hard gears coupled tightly, the following 
results are naturally predicted. When a torque works on a gear and this 
gear rotates in a direction, the rotational energy transmits to the other 
gear and both gears rotate with the coupling P). On the other hand, with the 
coupling T), such a transmission does not appear. For example we consider 
the case that a torque in the clockwise direction works on Gear 1. 
Due to the rotation of 
Gear 1, the torques in the anti-clockwise direction work on Gear 2 and 3. 
At the same time, however, the torque in the clockwise direction also work 
on Gear 2 and 3 by the interaction between them. Then, by this frustration, 
Gear 2 and 3 cannot rotate which means it is impossible to transmit the 
rotational energy to them. 

However, the different types of behaviors are expected if these gears are 
soft to deform largely. Then in the following, we study the responses of 
such a coupled gear system against the inputs for different softness 
of each gear. First of in this paper, we construct a model of the ideal 
coupled gears and focus on the responses of this system against a torque 
in a direction working on a gear. This model belongs to a specific case 
of coupled oscillator systems with frustrations\cite{kurapon,daido,awa}.
In this model, we obtain the opposite properties for the energy 
transmission between soft gears and hard gears: The soft gear system can 
realize the energy transmission to another gear only with the coupling T) 
while the hard gear system can realize the energy transmission only 
with the coupling P). Second, we show the responses of this system in which 
one of gears have contact with a high temperature heat bath and the other 
gears have contact with a $0$ temperature heat bath.
With the coupling T), the directional rotations appear in two gears 
with $0$ temperature heat bath.

Now, we construct a model of ideal coupled gear system as follows. 
First, we prepare a group of hard rods with unit length and set the edge 
of each rod at a rotation axis in two dimensional space. Second, we 
prepare particles between which repulsive forces work on, and put on the 
other edge of each rod. By these two steps, we can make a deformable gear. 
The coupled gear system is constructed as Fig. 1 by preparing the above 
mentioned gears and fixing the centers of them. In this model, the motions 
of particles indicate the motion of teeth of gears. In this paper, for 
simplicity, we consider the over-dumping limit cases as a motion of each 
particle. 

Kinetic equation for each particle is given by  
\begin{equation}
\dot{\bf r}^{i}_{j}= - \sum_{(i,j) \ne (i',j')} C^{i,i'}_{j,j'} \nabla_{{\bf r}^{i}_{j}} V(|{\bf r}^{i}_{j}-{\bf r}^{i'}_{j'}|) + {\bf F}^{i}_{j}.
\end{equation}
Here, ${\bf r}^{i}_{j}$ indicates a position of $j$th particle in $i$th 
gear in two dimensional space, and ${\bf F}^{i}_{j}$ indicates the torque 
working on $j$th particle in $i$th gear. In this paper, we employ 
$V(r)=1/r$ as the repulsive interaction potential between 
particles\cite{katasa}. 
$C^{i,i'}_{j,j'}$ gives the magnitude of the repulsive force working 
between $j$th particle in the $i$th gear and $j'$th particle in $i'$th 
gear. In the following, we set $C^{i,i'}_{j,j'}=1$ for $i \ne i'$ and 
$C^{i,i'}_{j,j'}=A$ for $i = i'$. Here, $A$ is a parameter indicating 
the hardness of each gear. In the followings, we consider the case that 
each gear has three teeth (rods). Qualitatively similar results as 
those observed in this paper are found if the number of teeth in each 
gear is more than three.

Now, the transmission and conversion of the energy by this coupled 
deformable gear system with the couplings P) and T) is studied for several 
hardness of gears $A$. Here, the distances between the centers of gears 
(the gear-gear distances) are given $L$.
First, we show the responses of coupled gears against a torque in a direction 
working on a gear. 

Figure 1(a) and 1(b) show the typical temporal evolutions of the tightly 
coupled hard gears (a) with the coupling P) and (b) with the coupling T), 
where $A=70$ and $L=1.8$ and torque in the clockwise direction works on 
the Gear 1 ($F^{1}_{j}=F$, $F^{2}_{j} = F^{3}_{j}=0$). Here, it is noted 
that we can observe only the periodic motions as the dynamical motions of 
the system in the presented situation independent of $A$ and $L$. 
(No quasi-periodic or chaotic motions appear.) In Fig. 1(a), both of two 
gears rotate with, respectively, the opposite directions. This means the 
rotational energy transmits from Gear 1 to Gear 2 in the system with the 
coupling P). On the other hand in Fig. 1(b), only Gear 1 rotates and each 
tooth of Gear 2 and 3 only oscillates in a restricted region. Thus, the 
rotational energy is not transmitted from Gear 1 to any other gears in the 
system with the coupling T). These results are consistent with the 
predictions for coupled hard gears. 

Now, we focus on the behavior of the coupled soft gears. Figure 1(c) and 
(d) show the typical temporal evolutions of the coupled soft gears (c) with 
the coupling P) and (d) with the coupling T) under the same situation as the 
previous cases except the hardness of each gear ($A=7$). As seen in them, the 
shape of gears are extremely different from those in hard gears systems 
because the teeth are pushed out from the inner area of the system by the 
repulsive forces between particles belonging to different gears. 
Then, two gears look not engaged. Thus, as shown in Fig. 2(a), only Gear 1 
rotates in the system with the coupling P), which means the rotational 
energy is not transmitted to Gear 2. On the other hand in the system with 
the coupling T), the rotational energy can be transmitted from Gear 1 to 
Gear 2 as shown in Fig. 1(d) where Gear 2 rotates 
in the anti-clockwise direction due to the clockwise rotation of Gear 1. 
In this case, each tooth in Gear 3 oscillate in a restricted area. 
(Because of the symmetry, the rotational energy transmits from Gear 1 to 
Gear 3 if the torque in the anti-clockwise direction works on Gear 1.) 

As previous, we obtained the clear evidences of the differences in the 
mechanical properties between the hard gear systems and the soft gear 
systems. In the following we study the detailed properties and mechanisms 
of the transmissions of soft gear system with the coupling T).

Figure 2 shows the time averaged angular velocity of each gear in (a) the 
hard gear system with $A=70$ and (b) the soft gear with $A=7$ with the 
coupling T) as a function of torque $F$ working on Gear 1 for $L=1.8$ and 
$L=2$. Here, the time averaged angular velocity is defined as 
$\lim_{\tau \to \infty} (RN(t+ \tau)-RN(t))/ \tau$ ($RN$ indicates the no. 
of rotations of the gear.), and the clockwise direction is defined as 
positive direction. As in Fig. 2(a), only Gear 1 rotates in the clockwise 
direction for $F$ larger than a critical value if each gear is hard. On 
the other hand, if each gear is soft, Gear 1 rotates in the clockwise 
direction and Gear 2 rotates in the anti-clockwise direction in certain 
range of $F$ in both cases with $L=1.8$ and $L=2$ (Fig. 2(b)). 
If $F$ is given much larger, only Gear 1 always rotates and rotational energy 
cannot transmit to any other gears independent of $A$.

Figure 2(c) shows the phase diagrams for the transmission properties of 
coupled gears as functions of the gear-gear distance $L$ and the hardness 
of each gear $A$. As found in Fig. 2(c), the energy transmission can realize 
only in the limited case with large enough $A$ and small enough $L$ in the 
case P). On the other hand with the coupling T), the energy transmission can 
realize in a wide range of $L$, if each gear is appropriately soft.

Now, we focus on the force balances of the coupled gears to clarify the 
mechanisms of the rotational energy transmissions between soft gears.
Figure 3 shows the snap shots of the time evolution of (a) hard gears 
$A=70$ and (b) soft gears $A=7$ with $L=1.8$. In Fig. 3(a), it is clear 
that the rotation direction of two teeth around the inner area of the 
triangle belonging to Gear2 and 3 and the direction of repulsive 
force working between them are close. Thus, Gear 2 and Gear 3 cannot 
rotate if each gear is hard. 

If each gear is soft, the relationships between 
the force directions and the rotational directions change drastically 
as in Fig. 3(b). In this case, following force relations appear against the 
torque in the clockwise direction working on Gear 1. I) The tooth of 
Gear 1 at the inner area of the triangle pushes the left tooth of Gear 3. 
II) The right tooth of Gear 3, which is pushed by the left tooth through a 
tooth between the left and right teeth, pushes the bottom tooth in Gear 2.
III) The left tooth of Gear 2, which is pushed by the bottom tooth through a 
tooth between the left and bottom teeth, pushes the top tooth of Gear 1. 
IV) The direction of the force between the left tooth in Gear 2 and the 
top tooth in Gear 1 is not close to the opposite direction of the movement 
of the left tooth in Gear 2 but close to that of the top tooth in Gear 1. 
Here, The fact I) induces the fact II), and 
II) induces III). By the facts III) and IV), the left tooth in Gear 2 can 
move to the inner area of the triangle, and after this, the top tooth in 
Gear 1 moves to the inner area. Thus, the rotational energy can transmit 
to Gear 2 (Gear 3) from Gear 1 if the torque in the clockwise 
(anti-clockwise) direction works on Gear 1.

Next, we focus on the responses of the coupled soft gears with the 
coupling T) against the random torque working on each particle in Gear 1.
Here, we consider the situation where Gear 1 has contact with a heat bath with 
temperature $\theta$, and the Gear 2 and Gear 3 have contact with a heat bath 
with temperature $0$.
In this situation, as in the followings, the directional 
motions appear in Gear 2 and Gear 3 in the certain range of $\theta$, 
where these two gears rotate in opposite directions with each other.

Figure 4(a) and 4(b) show the typical temporal evolutions of no. of 
rotations of one of particles in each gear for (a) $\theta=0.125$ and (b) 
$\theta=0.5$, and Fig. 4(c) shows the time averaged angular velocity of 
Gear 2 as a function of $\theta$ for $A=7$ and $L=1.8$. In several 
$\theta$, while the motion of Gear 1 looks like Brownian motion, 
Gear 2 and Gear 3 move in opposite directions in average with 
each other. 
Moreover, the directions of each gears 
rotation change depending on $\theta$, where Gear 2 (3) rotates in the 
anti-clockwise (clockwise) direction for small $\theta$ or the clockwise 
(anti-clockwise) direction for large $\theta$. (For much larger $\theta$, 
the rotations of Gear 2 and Gear 3 becomes $0$.) Thus, this system may play 
the roles of not only rectifier but also the noise strength detector. 

Here, we focus on the detailed motion of the coupled soft gears due to 
the energy conversions. When $\theta$ is small, the magnitude of the 
torque working on Gear 1 is not so large in average. In such cases, as in 
Fig 2(b), Gear 2 or 3 can rotate with Gear 1. Then, Gear 2 (3) can rotate 
in the anti-clockwise (clockwise) direction when Gear 1 happens to 
rotate in the clockwise (anti-clockwise) direction. Thus, Gear 2 (3) 
rotates in the anti-clockwise (clockwise) direction in average.

As in Fig 2(b), Gear 2 or 3 cannot rotate with Gear 1 if $F$ is much larger. 
Similarly, if $\theta$ becomes large, the detailed motions of Gear 1 
affects little to those of Gear 2 and Gear 3 while the random torques 
continue to work on Gear 2 and Gear 3 by the heat conduction from Gear 1. 
Then, with the certain $\theta$, the situation where the interaction 
between Gear 2 and 3 is still effective may realize.
In such cases, when Gear 2 happens to move 
in the clockwise direction, Gear 3 can also rotate in the anti-clockwise 
direction, or when Gear 3 happens to move in the anti-clockwise direction, 
Gear 2 can also rotate in the clockwise direction. In this case, the 
timing of the rotations of Gear 2 and 3 correlate with each other, which is 
consistent to as seen in Fig 4(b). Thus, Gear 2 and Gear 3 rotate in the 
clockwise and anti-clock directions, respectively, in average for large 
$\theta$.

In this paper, the transmission and the conversion of the rotational energy 
by the coupled deformable gears are investigated. If each gear is soft, 
following dynamics are observed. A) The coupled gears can realize the 
transmission of rotational energy from one to the other only when three 
gears are coupled where they form a triangle, while the coupled hard gears 
can realize the  transmission only when two gears are tightly coupled. 
B) If one of gears have contact with a high temperature heat bath and the 
other gears have contact with a $0$ temperature heat bath with the coupling 
T), the directional rotations appear in two gears with $0$ temperature heat 
bath. These results give an explicit evidence of that the performances and 
mechanical properties of hard machines and soft machines are very different. 

For each case, there are some kinds of merits and demerits. In general, the 
tightly coupled two hard gears are expected to realize the transmissions in 
wide range of the rotational energy. However, if the distance between two 
gears becomes a little long, these systems cannot realize the transmissions 
suddenly which means this system is weak for fluctuations. 
On the other hand, in the coupled soft gears forming a triangle, the range 
of the rotational energy which can be transmitted from one to others is 
limited. In this case, however, the such transmissions can occur almost 
independently of the distances between gears. This fact expects that the 
coupled soft gears realize such a transmission robustly against the 
fluctuations. This fact implies the soft gears are useful for making micro or 
nano scale machines. This fact may also imply that the softness of biological 
molecular machines are suitable to work in large influences of fluctuations.

Analytical study of the presented system, as well as the discoveries of 
possible dynamics and realization of functions of more varieties in models 
of soft machines should be important as future issues.  

The author is grateful to M. Sano, K. Kaneko and N. Kataoka for useful 
discussions. This research was supported in part by a Grant-in-Aid for 
JSPS Fellows (10039).

\begin{figure}
\begin{center}
\includegraphics[width=8.8cm]{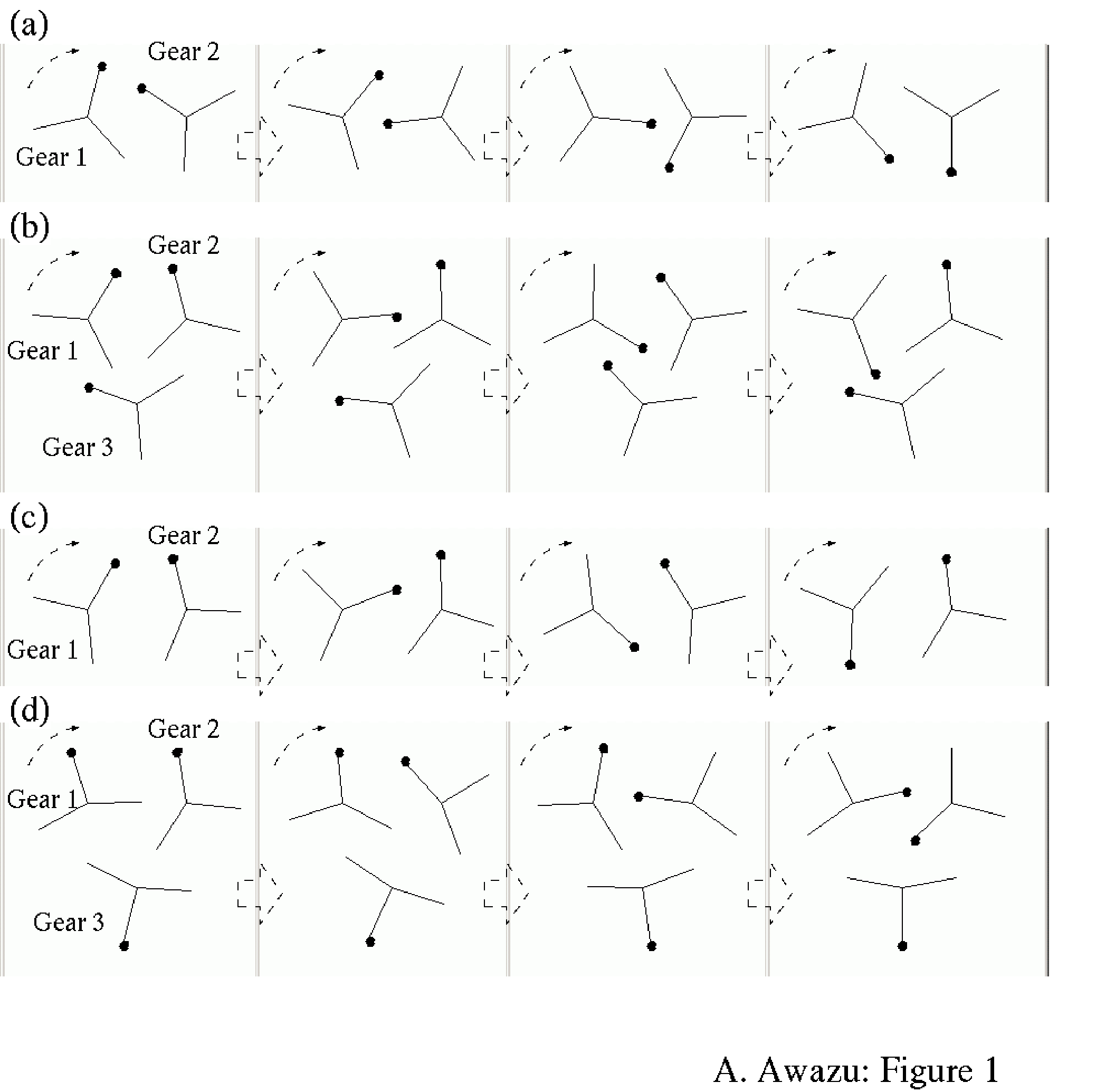}
\end{center}
\caption{Typical temporal evolutions of the system consists of 
(a) $A=70$ with the coupling P), (b) $A=70$ with the coupling T), 
(c) $A=7$ with the coupling P) and (d) $A=7$ with the coupling T) (L=1.8). 
Dashed arrows indicate the direction of the torque $F$, and circles are 
just markers.
}
\end{figure}

\begin{figure}
\begin{center}
\includegraphics[width=8.0cm]{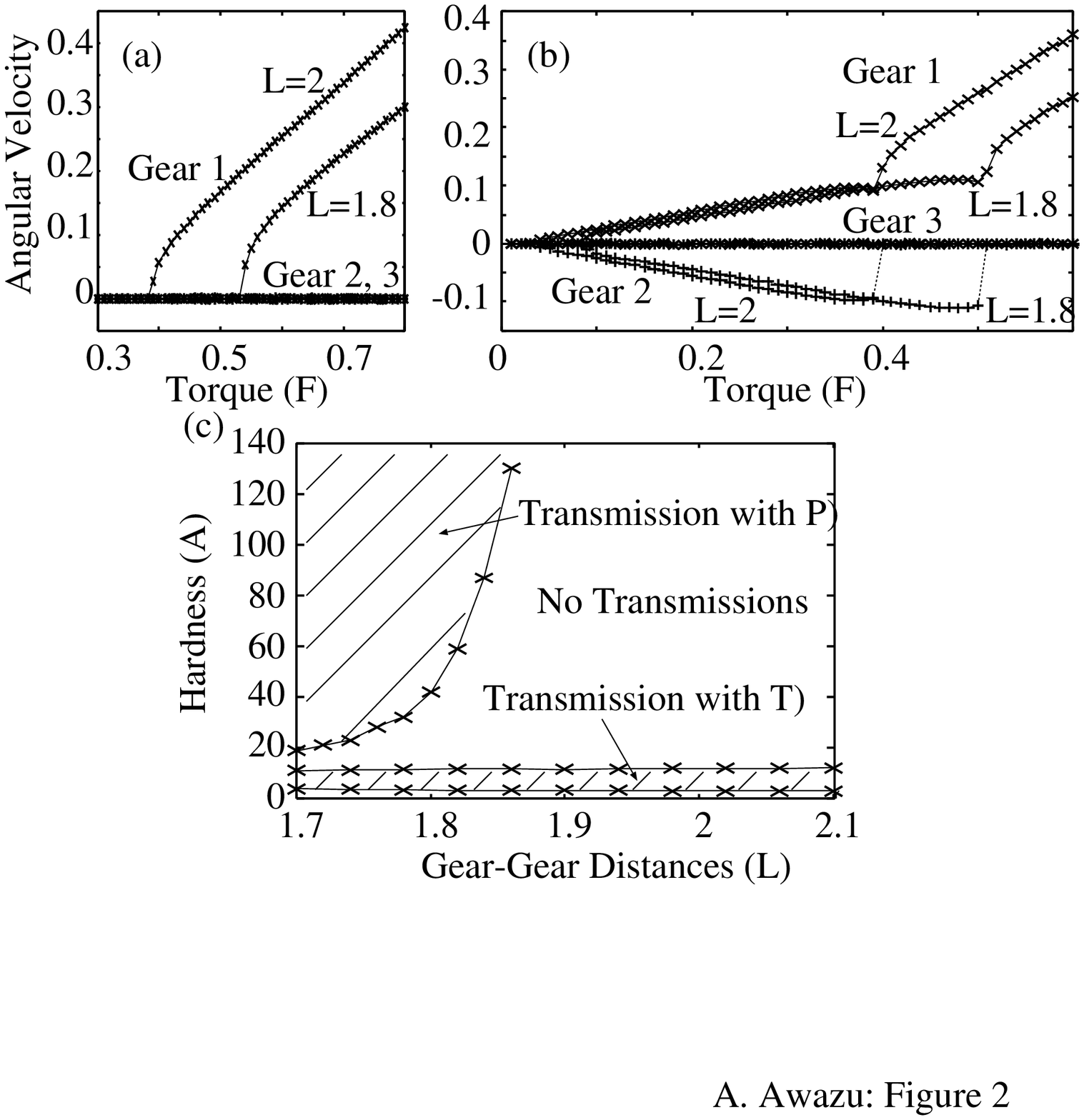}
\end{center}
\caption{Time averaged angular velocity of each gear in the systems with
(a)$A=70$ and (b)$A=7$ for $L=1.8$ and $L=2$. (c) Phase diagram of the system 
for the rotational energy transmissions.
}
\end{figure}

\begin{figure}
\begin{center}
\includegraphics[width=8.0cm]{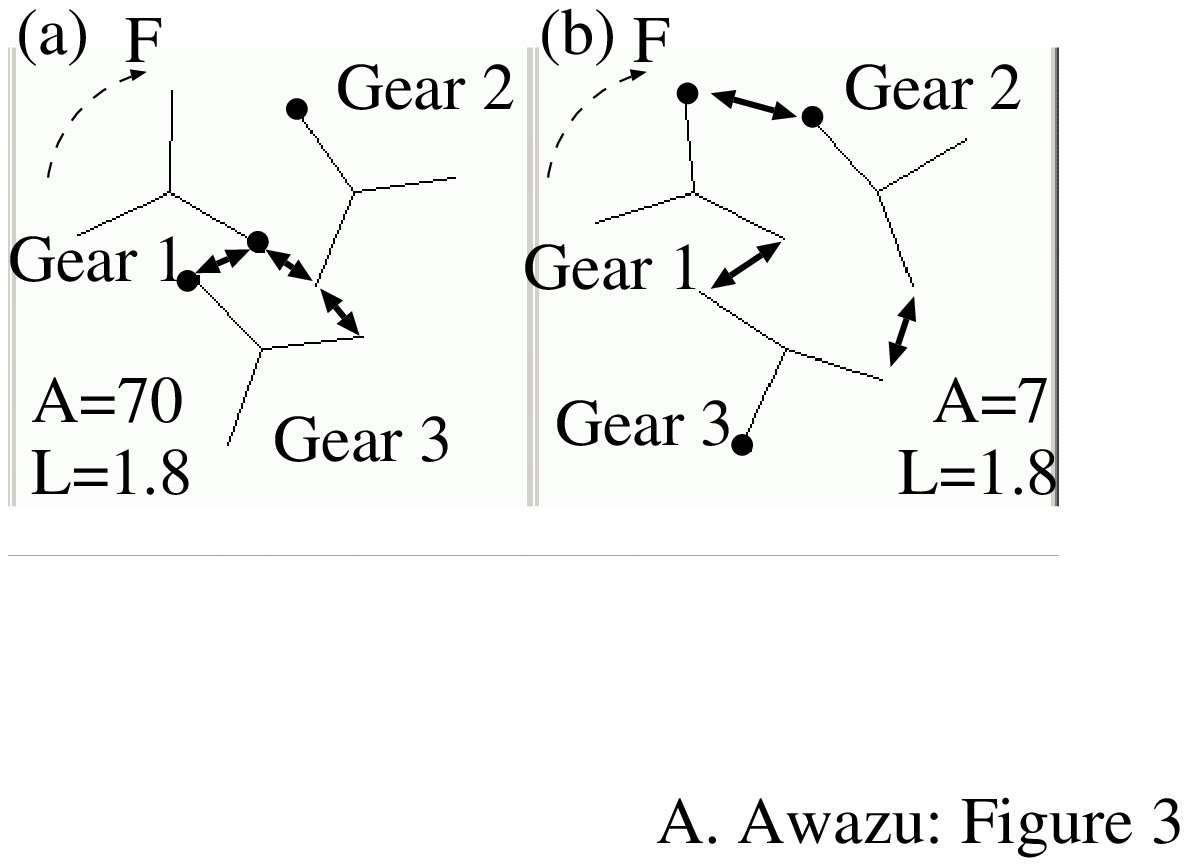}
\end{center}
\caption{Force Relationships between gears.}
\end{figure}

\begin{figure}
\begin{center}
\includegraphics[width=8.0cm]{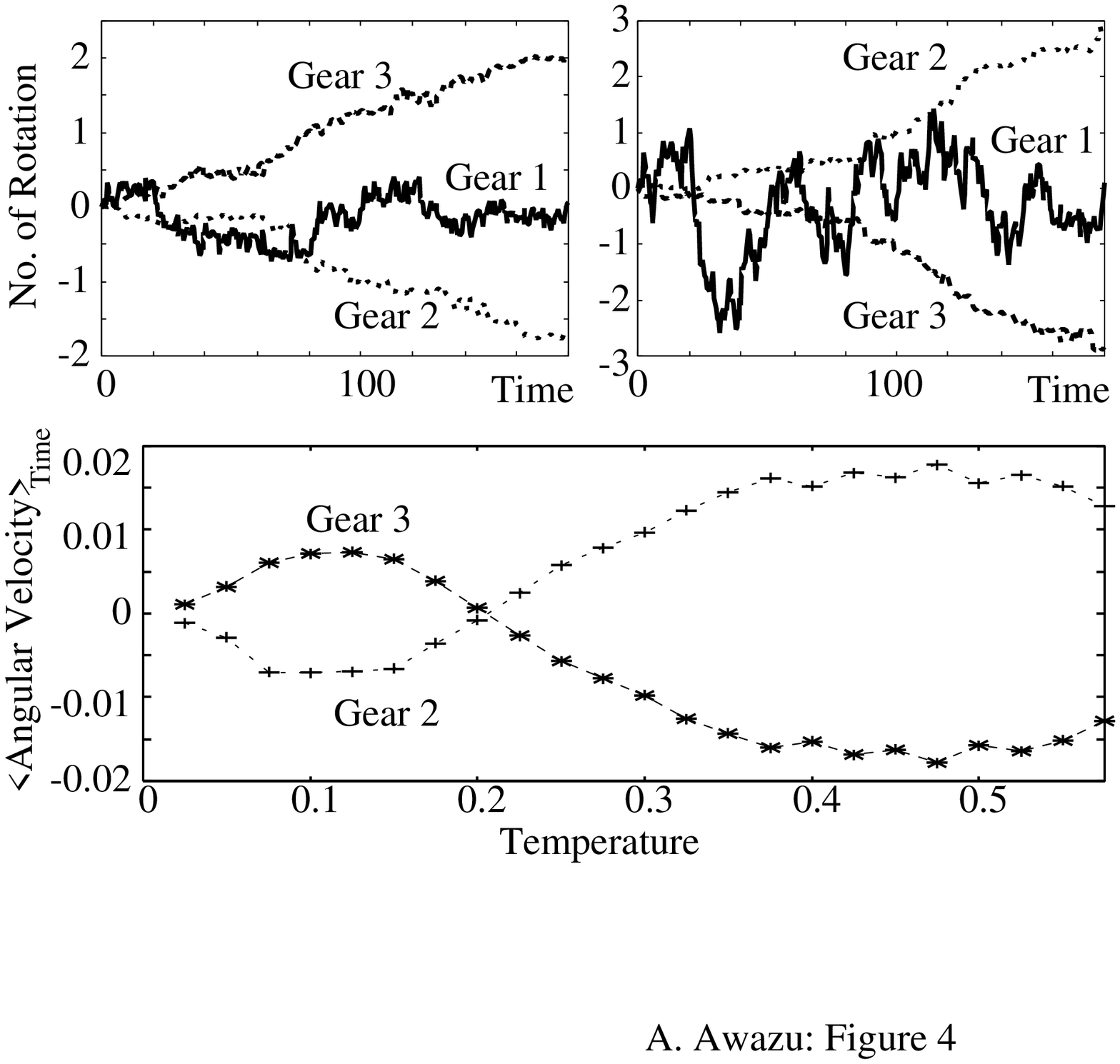}
\end{center}
\caption{Typical temporal evolutions of no. of rotations of a particle in 
each gear for (a)$\theta=0.125$ and (b)$\theta=0.5$, and (c) 
Time averaged angular velocity of Gear 2 and Gear 3 as a function of 
$\theta$ for $A=7$ and $L=1.8$.}
\end{figure}


\begin{thebibliography}{9}
\bibitem{katasa} 
In cases $V(r)=(1/r)^{n}$ with $n>1$, we can obtain the similar 
results as obtained in this paper.
\end{thebibliography}
\end{document}